\documentclass{rnaastex}

\begin{document}

\title{ASASSN-18di: discovery of a $\Delta V \sim 10$ flare on a mid-M dwarf}






\correspondingauthor{R. Rodr\'iguez}
\email{rodriguezmartinez.2@osu.edu}

\author[0000-0002-6244-477X]{R. Rodr\'iguez}
\affiliation{Department of Astronomy, The Ohio State University, 140 West 18th Avenue, Columbus, OH 43210, USA}

\author{S. J. Schmidt}
\affiliation{Leibniz-Institut f\"ur Astrophysik,
An der Sternwarte 16
14482 Potsdam, Germany}

\author{T. Jayasinghe}
\affiliation{Department of Astronomy, The Ohio State University, 140 West 18th Avenue, Columbus, OH 43210, USA}

\author{K. Z.  Stanek}
\affiliation{Department of Astronomy, The Ohio State University, 140 West 18th Avenue, Columbus, OH 43210, USA}
\affiliation{Center for Cosmology and Astroparticle Physics, The Ohio
State University, 191 W. Woodruff Avenue, Columbus, OH
43210}

\author{J. L. Prieto}
\affiliation{N\'ucleo de Astronom\'ia de la Facultad de Ingenier\'ia y Ciencias,
Universidad Diego Portales, Av. Ej\'ercito 441, Santiago,
Chile}
\affiliation{Millennium Institute of Astrophysics, Santiago, Chile}

\author{B. Shappee}
\affiliation{Institute for Astronomy, University of Hawai'i, 2680 Woodlawn Drive, Honolulu, HI 96822, USA}

\author{C.S. Kochanek}
\affiliation{Department of Astronomy, The Ohio State University, 140 West 18th Avenue, Columbus, OH 43210, USA}

\author{Todd A. Thompson}
\affiliation{Department of Astronomy, The Ohio State University, 140 West 18th Avenue, Columbus, OH 43210, USA}

\author{J. Shields}
\affiliation{Department of Astronomy, The Ohio State University, 140 West 18th Avenue, Columbus, OH 43210, USA}

\author{T.W.-S. Holoien}
\affiliation{The Observatories of the Carnegie Institution for Science, 813 Santa Barbara Street, Pasadena, CA, 91101, USA}

\author{D. Bersier}
\affiliation{Astrophysics Research Institute, Liverpool John Moores University, 146 Brownlow Hill, Liverpool L3 5RF, UK}

\author{J. Brimacombe}
\affiliation{Coral Towers Observatory, Cairns, Queensland 4870, Australia}


\section{} 

We report the detection of ASASSN-18di, a powerful white-light superflare on a previously undiscovered, mid-type
M dwarf located at R.A. 9:58:30.51 and decl. 14:55:39.0. This flare was discovered by the All-Sky Automated Survey for SuperNovae (ASAS-SN, \citealt{2014ApJ...788...48S}) and represents one of the largest flares ever observed on a cool dwarf. The three detections of the flare occurred on UT 2018 February 20th, with the first and last exposures separated by $\Delta t \sim 5$ mins. We do not find evidence of additional flares at the location of this transient in ASAS-SN archival images from UT 2012 October to 2018 March, only upper limits at $V \sim 17.5$.  This is consistent with the low occurrence rate of $\sim 1/10 \,\, {\rm yr^{-1}}$ expected for $E\sim 10^{36}$ ergs flares on mid-M dwarfs based on \citet{2016ApJ...829L..31D}. This discovery adds to the number of dramatic flares detected on cool dwarfs.

ASAS-SN is particularly well suited for the detection of white-light flares, as it is monitoring the entire sky with daily cadence down to $g\sim 18$, allowing us to detect $E\sim 10^{34}$ ergs flares on cool dwarfs out to $100$ pc and $E\sim 10^{36}$ ergs flares out to $1000$ pc. In fact, ASAS-SN has already discovered many stellar flares, including two strong ones (\citealt{2014ApJ..781...24S, 2014ApJ..828...22S}), from an M8 and L1 dwarf, respectively. These stars were bright enough for the authors to spectroscopically confirm and characterize the host stars while constraining flare energetics using classical-flare models. Unfortunately, the source presented here is too faint for follow-up characterization without the investment of significant telescope time. In addition, it is unlikely that we will observe another flare of such amplitude, given their low occurrence rate.

We complement the ASAS-SN photometry given in Table 1; \citep{2017PASP..129j4502K}, by cross-matching sources within $3\arcsec$ of our ASAS-SN position, in both the DECaLS and \textit{WISE} \citep{2014AJ...140...1868W} databases. From the former catalog, we obtain fluxes for the $r$ and $z$ bands corresponding to $r= 23.2$ and $z=21.1$ mag. The source is undetected in the $u$, $g$, and $i$ bands. The $r-z$ colors are consistent with a spectral type of M4, based on the color-spectral-type relations from \citet{2011AJ....141...97W}.
However, the \textit{WISE} \textit{W1-W2} colors are slightly redder than expected for a mid M dwarf, based on the empirical relationships for ultracool dwarfs in \citet{2015AJ....149..158S}. To estimate a quiescent $V$ magnitude, we employ the color-magnitude transformations from Schmidt et al. (2018, in prep), and derive a value of $V \simeq 24.0$ mag. This implies a flare magnitude of $\Delta V \simeq 9.8$ mag. Given the large estimated flare amplitude $\Delta V$, it is unlikely that we are observing this flare during either its rapid impulse or decay phase and we assume that the flare was detected near its peak. We derive an absolute magnitude of the star of $M_{r}=11.4$ from \citet{2010AJ....139.2679B}, leading to an approximate distance of $d \sim 2.2$ kpc. We place a constraint on the total flare energy of $E_{V} \simeq (4.1\pm 2.2)\times 10^{36}$ ergs, based on Schmidt et al. (2018, in prep), thus placing ASASSN-18di amongst the strongest flares documented on M dwarfs \citep{2016ApJ...829...23D}.

\newpage

Today it is widely known that M-type and ultracool stars can display high levels of magnetic activity, and numerous large flares have been discovered originating from cool dwarfs. However, in order to study the flaring rates, magnetic dynamos and interiors of these stars down to the fully convective and substellar regimes, we need more detections of flare events. This is especially true for mid-to-late M dwarfs and stars at the M/L transition, where the literature is most sparse. Recent studies also indicate that magnetic activity plays an especially important role on the detection and habitability of extrasolar planets around active stars \citep{2018arXiv180302338T,2010AsBio..10..751S}. The upcoming NASA TESS mission \citep{2015JATIS...1a4003R} will offer a unique opportunity to enrich our catalog of well-characterized flares, as it will take high precision photometry of many M and ultracool dwarfs at a higher cadence than ASAS-SN. Furthermore, because many of these targets are bright and nearby, they will offer promising chances for follow-up characterization with JWST.


\begin{table*}
\caption{Photometry}
\centering
\begin{tabular}{ccccccc}
\hline\hline
Photometric Band & Source & Mag & Mag Err & Flux (mJy) & Flux Err & JD \\
\hline
$V$ & ASAS-SN & 14.24 & 0.03 & 7.66 &  0.24 & 2458169.60387 \\
$V$ & ASAS-SN & 14.27 & 0.03 & 7.51 &  0.25 & 2458169.60515 \\
$V$ & ASAS-SN & 14.23 & 0.04 & 7.76 &  0.24 & 2458169.60644  \\
\hline
$r$ & DECaLS & 23.2 & 0.15 & 0.49 & 0.07 \\
$z$ & DECaLS & 21.1 & 0.04 & 3.38 & 0.14\\
$W1$ & $WISE$ & 16.64 & 0.09 &\\
$W2$ & $WISE$ & 16.38 & 0.28 & \\
\hline

\end{tabular}
\label{tab:pointings}
\bigskip
\end{table*}



\acknowledgments

This project used data obtained with the Dark Energy Camera (DECam), which was constructed by the Dark Energy Survey (DES) collaboration. 



\begin{thebibliography}{}


\bibitem[Bochanski et al.(2010)]{2010AJ....139.2679B} Bochanski, J.~J., Hawley, S.~L., Covey, K.~R., et al.\ 2010, \aj, 139, 2679-2699 

\bibitem[Davenport et al.(2016)]{2016ApJ...829L..31D} Davenport, J.~R.~A., Kipping, D.~M., Sasselov, D., Matthews, J.~M., \& Cameron, C.\ 2016, \apjl, 829, L31 


\bibitem[Davenport(2016)]{2016ApJ...829...23D} Davenport, J.~R.~A.\ 2016, \apj, 829, 23 


\bibitem[Kochanek et al.(2017)]{2017PASP..129j4502K} Kochanek, C.~S., Shappee, B.~J., Stanek, K.~Z., et al.\ 2017, \pasp, 129, 104502 


\bibitem[Ricker et al.(2015)]{2015JATIS...1a4003R} Ricker, G.~R., Winn, J.~N., Vanderspek, R., et al.\ 2015, Journal of Astronomical Telescopes, Instruments, and Systems, 1, 014003 

\bibitem[Schmidt et al.(2015)]{2015AJ....149..158S} Schmidt, S.~J., Hawley, S.~L., West, A.~A., et al.\ 2015, \aj, 149, 158 


\bibitem[Schmidt et al.(2014)]{2014ApJ..781...24S} Schmidt, S.~J., J., Prieto, J.~L., Stanek, K. ~Z. et al.\ 2014, \apj, 781, 24 

\bibitem[Schmidt et al.(2016)]{2014ApJ..828...22S} Schmidt, S.~J., J., Shappee, B.~J., Gagn\'e, J., Stanek, K. ~Z. et al.\ 2016, \apj, 828, 22 

\bibitem[Segura et al.(2010)]{2010AsBio..10..751S} Segura, A., Walkowicz, L.~M., Meadows, V., Kasting, J., \& Hawley, S.\ 2010, Astrobiology, 10, 751 


\bibitem[Shappee et al.(2014)]{2014ApJ...788...48S} Shappee, B.~J., Prieto, J.~L., Grupe, D., et al.\ 2014, \apj, 788, 48

\bibitem[Tal-Or et al.(2018)]{2018arXiv180302338T} Tal-Or, L., Zechmeister, M., Reiners, A., et al.\ 2018, arXiv:1803.02338 



\bibitem[West et al.(2011)]{2011AJ....141...97W} West, A.~A., Morgan, D.~P., Bochanski, J.~J., et al.\ 2011, \aj, 141, 97 


\bibitem[Wright et al.(2010)]{2014AJ...140...1868W} Wright, E.~L., Eisenhardt, Peter R.M., Mainzer, A.K., et al.\ 2010, \aj, 140, 1868






\end{thebibliography}
\end{document}